\documentclass[12pt]{article}
\usepackage{epsfig}

\def\pl#1#2#3{Phys.~Lett.~{\bf B {#1}} (19{#2}) #3}

\def\pr#1#2#3{Phys.~Rev.~{\bf D {#1}} (19{#2}) #3}

\begin{document}
\date{}

\title{ 
{\normalsize
\mbox{ }\hfill
\begin{minipage}{3cm}   
DESY 01-046\\
OUTP-01-19-P
\end{minipage}}\\
\vspace{3cm}
\bf Spectator Processes\\ and Baryogenesis}
\author{W.~Buchm\"uller\\
{\it Deutsches Elektronen-Synchrotron DESY, 22603 Hamburg, Germany}\\[5ex]
M.~Pl\"umacher\\
{\it Theoretical Physics, University of Oxford, 1 Keble Road,}\\
{\it Oxford, OX1 3NP, United Kingdom}
}
\maketitle

\thispagestyle{empty}

\vspace{2cm}
\begin{abstract}
\noindent
Spectator processes which are in thermal equilibrium during the period of
baryogenesis influence the final baryon asymmetry. We study this effect
quantitatively for thermal leptogenesis where we find a suppression by
a factor {\cal O}(1).
\end{abstract}

\newpage

The cosmological baryon asymmetry is naturally explained by thermal
leptogenesis, i.e. the out-of-equilibrium decays of heavy Majorana 
neutrinos \cite{fy86}. This mechanism has been studied in detail by several
groups \cite{bp00}, and it has been shown that the observed
baryon asymmetry can be obtained over a wide range of parameters which 
are consistent with the observed properties of light neutrinos.

In a complete analysis decays, inverse decays and various scattering
processes in the primordial plasma
have to be taken into account in order to get a reliable
estimate of the generated baryon asymmetry \cite{lut92,plu97}.
In all previous analyses only the chemical potential $\mu_L$ of the standard 
model (SM) leptons was treated as a dynamical variable during the process of
leptogenesis. All other chemical potentials, and also the baryon asymmetry,
were then obtained from $\mu_L$ assuming thermal equilibrium after the 
period of leptogenesis.

However, this picture is incorrect. The duration $\tau_L$ of leptogenesis, 
which is driven by the out-of-equilibrium decays of heavy Majorana 
neutrinos, is larger than the Hubble time, $\tau_L > 1/H$. On the other
hand, many processes in the plasma, in particular the sphaleron processes, 
are in thermal equilibrium, i.e.\ $\tau_i = 1/\Gamma_i < 1/H$. Hence, many 
spectator processes in the plasma are faster than leptogenesis. As a 
consequence the chemical potentials of particles in thermal equilibrium 
are changed already during the process of leptogenesis. In the following 
we study the effect of these spectator processes on the final baryon 
asymmetry in the framework of the SM with additional heavy Majorana 
neutrinos $N_j$.\\

\noindent
\textbf{Boltzmann equations}\\

\noindent
The most important processes that have to be considered are decays and
inverse decays, whose reaction densities will be denoted by
$\gamma_{D,j}$ in the following, top-quark neutrino scatterings
mediated by a SM Higgs in the $s$- or $t$-channel, denoted by
$\gamma_{\phi,s}^j$ and $\gamma_{\phi,t}^j$, and $\Delta L=2$
processes mediated by heavy right-handed neutrinos in the $s$- or
$t$-channel, denoted by $\gamma_N$ and $\gamma_{N,t}$ (cf.~\cite{bp00}). 
Hence, in
addition to the lepton number $Y_{\rm L}$ we also have to consider the 
chemical potentials of the Higgs doublet $\phi$, of the quark doublets
$Q$ and of the right-handed top quarks $t$. Since the Yukawa couplings 
of up and charm quarks are significantly smaller they do not have to
be considered here. We will use the following notation for the
particle number asymmetries:
\begin{equation}
Y_{\rm H}={n_{\phi}-n_{\bar{\phi}}\over s}, \qquad
Y_{\rm Q}={n_Q-n_{\bar{Q}}\over s} \quad\mbox{ and } \quad
Y_{\rm T}={n_t-n_{\bar{t}}\over s}\;,
\end{equation}
where $s$ denotes the entropy density of the universe and $n_i$ are the
number densities of the corresponding particles.

Taking these chemical potentials into consideration one obtains after a
straightforward calculation the
following set of Boltzmann equations for the evolution of the number
of heavy Majorana neutrinos $Y_{N_j}$ and the lepton asymmetry
$Y_{\rm L}$:
\begin{eqnarray}
  {{\rm d}Y_{N_j}\over{\rm d}z} &=& - {z\over sH(M_1)} 
    \left({Y_{N_j}\over Y_{N_j}^{\rm eq}}-1\right) 
    \left(\gamma_{D,j}+2\gamma_{\phi,s}^1+4\gamma_{\phi,t}^1\right)
    \label{Boltzmann_eq1}\\[1ex]
  {{\rm d}Y_{\rm L}\over{\rm d}z} &=&  - {z\over sH(M_1)} 
    \left\{
      \sum\limits_j \left[
        {1\over2}\left({Y_{\rm L}\over Y_l^{\rm eq}}
                       +{Y_{\rm H}\over Y_{\phi}^{\rm eq}}\right)
        -\varepsilon_j\left({Y_{N_j}\over Y_{N_j}^{\rm eq}}-1\right)
        \right]\gamma_{D,j} \right.\nonumber\label{Boltzmann_eq2}\\[1ex]
        &&\hspace{1cm}+\; 2\left({Y_{\rm L}\over Y_l^{\rm eq}}
          +{Y_{\rm H}\over Y_{\phi}^{\rm eq}}\right)\left(
           \gamma_N+\gamma_{N,t}\right)\nonumber\\[1ex]
        &&\hspace{1cm}\left.+\sum\limits_j \left[
          \left({Y_{N_j}\over Y_{N_j}^{\rm eq}}
            {Y_{\rm L}\over Y_l^{\rm eq}}
            +{Y_{\rm T}\over Y_t^{\rm eq}}
            -{Y_{\rm Q}\over Y_q^{\rm eq}}\right)\gamma_{\phi,s}^j
            \right.\right.\nonumber\\[1ex]
        &&\hspace{1cm}\left.\left. +\left(2{Y_{\rm L}\over Y_{\rm L}^{\rm eq}}
            +\left({Y_{\rm T}\over Y_t^{\rm eq}}
            -{Y_{\rm Q}\over Y_q^{\rm eq}}\right)
            \left({Y_{N_j}\over Y_{N_j}^{\rm eq}}+1\right)
          \right)\gamma_{\phi,t}^j\right]
      \right\}\;.
\end{eqnarray}
Here $Y_{N_j}^{\rm eq}$, $Y_l^{\rm eq}$, $Y_{\phi}^{\rm eq}$,
$Y_q^{\rm eq}$, and $Y_t^{\rm eq}$ are the ratios of number density
and entropy density for Majorana neutrinos,
lepton doublets, Higgs doublet, quarks doublets and right-handed
top-quarks, respectively. Further, we have assumed that all chemical
potentials and CP asymmetries are small, and we have linearized the
Boltzmann equations in CP and particle asymmetries.
Setting all particle number asymmetries, except 
$Y_{\rm L}$, to zero one recovers the Boltzmann equations that have
been studied previously \cite{lut92,plu97}.

It is interesting that the Boltzmann equation
(\ref{Boltzmann_eq1}) for Majorana neutrinos is not modified to
leading order in the chemical potentials. Similarly, the terms
linear in $\epsilon_j$, which are
responsible for the generation of a lepton asymmetry in
eq.~(\ref{Boltzmann_eq2}), remain unchanged whereas the washout terms
are enhanced. Indeed, decays of heavy
Majorana neutrinos will create identical particle number asymmetries
in lepton and Higgs doublets. Scattering processes in thermal equilibrium
can change the ratio of $Y_L$ and $Y_H$ but not the sign.
Further, since the top
quark Yukawa coupling is in thermal equilibrium, the particle number 
asymmetries of quarks are given by the Higgs doublet asymmetry,
\begin{equation}
  {Y_{\rm T}\over Y_t^{\rm eq}}-{Y_{\rm Q}\over Y_q^{\rm eq}}
  = {Y_{\rm H}\over Y_{\phi}^{\rm eq}}\;.
\end{equation}
Hence, since $Y_L$ and $Y_H$ have the same sign, every washout term
in eq.~(\ref{Boltzmann_eq2}) is enhanced.

\begin{figure}[t]
   \centerline{\epsfig{file=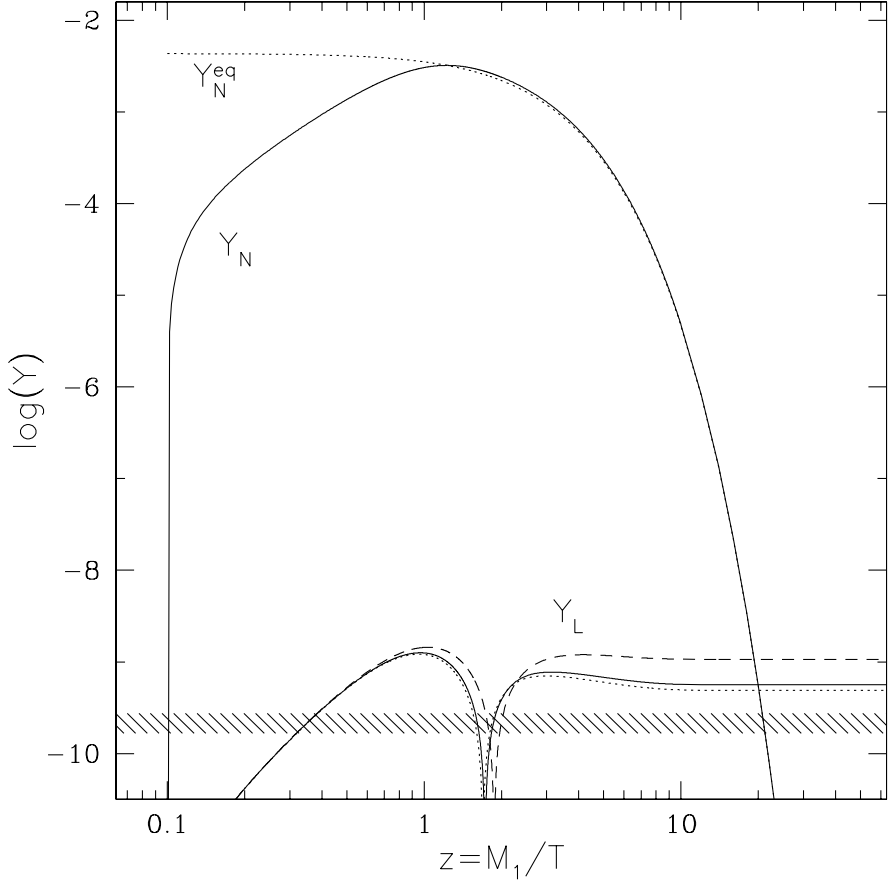,width=12cm}}
   \caption{{\it Solutions of the Boltzmann equations. Neglecting the
     chemical potentials of quarks and Higgs fields gives the
     dashed line for the lepton asymmetry $Y_{\rm L}$, whereas 
     the solid and dotted lines correspond
     to setting $N_F=1$ or $N_F=3$ in the Boltzmann 
     eq.~(\ref{Boltzmann_eq3}).}\label{figure}}
\end{figure}

Due to SM
gauge, Yukawa and non-perturbative sphaleron interactions not all
chemical potentials are independent. In the simplest case, when all 
SM interactions are assumed to be in thermal equilibrium, all chemical 
potentials can be expressed in terms of a single chemical potential,
which can be chosen to be $\mu_l$ (cf.~\cite{bp00}). 
In this case the Boltzmann equation (\ref{Boltzmann_eq2}) for the
lepton number takes the following form
\begin{eqnarray}
  {{\rm d}Y_{\rm L}\over{\rm d}z} &=&  - {z\over sH(M_1)} 
    \left\{
      \sum\limits_j \left[
        {1\over2}{14N_F+3\over6N_F+3}{Y_{\rm L}\over Y_l^{\rm eq}}
        -\varepsilon_j\left({Y_{N_j}\over Y_{N_j}^{\rm eq}}-1\right)
        \right]\gamma_{D,j} \right.\label{Boltzmann_eq3}\\[1ex]
        &&+ 2{14N_F+3\over6N_F+3}{Y_{\rm L}\over Y_l^{\rm eq}}
           \left(\gamma_N+\gamma_{N,t}\right)\nonumber\\[1ex]
        &&\left.+{Y_{\rm L}\over Y_l^{\rm eq}}
          \sum\limits_j \left[
          \left({Y_{N_j}\over Y_{N_j}^{\rm eq}}
            +{4N_F\over6N_F+3}\right)\gamma_{\phi,s}^j
          +\left(2\ {8N_F+3\over6N_F+3}+
            {4N_F\over6N_F+3}{Y_{N_j}\over Y_{N_j}^{\rm eq}}
           \right)\gamma_{\phi,t}^j\right]
      \right\}\;.\nonumber
\end{eqnarray}
Hence, taking the chemical potentials of all particles in the plasma into
account enhances the washout processes by a factor ${\cal O}(1)$.

\noindent
\textbf{Numerical results}\\

\noindent
The structure of couplings between right-handed Majorana neutrinos and 
SM fields is model dependent. However, generically the lightest
right-handed neutrino, which is responsible for the final lepton
asymmetry, is coupled predominantly to SM quarks and leptons of the
third generation. Hence, as a first approximation, we can neglect the
first two generations of SM quarks and leptons and numerically solve
the Boltzmann eqs.~(\ref{Boltzmann_eq1}) and (\ref{Boltzmann_eq3})
assuming one generation of light fermions with Yukawa interactions in
thermal equilibrium, i.e.\ setting $N_F=1$ in eq.~(\ref{Boltzmann_eq3}).

The result is shown in fig.~\ref{figure}, where we have used a
parametrization of neutrino masses and Yukawa couplings based on a
$SU(5)\times U(1)_F$ family symmetry described in 
ref.~\cite{by99}. We have superimposed the result obtained when all
chemical potentials, except lepton number, are neglected, which formally
corresponds to setting $N_F=0$ in eq.~(\ref{Boltzmann_eq3}).
To illustrate the effect of additional fermions we have also included the
results obtained when setting $N_F=3$. The final lepton
number generated in these three cases is
\begin{equation}
  Y_{\rm L}=\left\{\begin{array}{rl}
            1.1\cdot10^{-9} &\quad\mbox{for}\quad N_F=0\;,\\[1ex]
            5.6\cdot10^{-10}&\quad\mbox{for}\quad N_F=1\;,\\[1ex]
            4.9\cdot10^{-10}&\quad\mbox{for}\quad N_F=3\;.
            \end{array}\right.
\end{equation}
Hence, spectator processes in the standard model plasma reduce the
final lepton asymmetry by a factor of about 2.
The final baryon asymmetry is given by 
$Y_{\rm B} = - (8N_F+4)/(14N_F+9)Y_{\rm L}$.

We conclude that spectator processes generically influence washout 
effects and thereby 
modify the generated baryon asymmetry by a factor ${\cal O}(1)$.\\

\noindent
\textbf{Acknowledgements:}
M.P.~was supported by the EU network ``Supersymmetry and the
Early Universe'' under contract no.~HPRN-CT-2000-00152.

\end{document}